\RequirePackage{fix-cm}
\documentclass[smallextended]{svjour3}       
\smartqed  
\usepackage{graphicx}

\usepackage[utf8]{inputenc}
\usepackage[english]{babel}
\usepackage[sorting=none]{biblatex}
\usepackage{floatrow}
\usepackage{bm}

\begin{document}

\title{Generalized parton distributions of light nuclei
}


\author{Sara Fucini        \and
        Matteo Rinaldi
        \and Sergio Scopetta 
}


\institute{S. Fucini \at
              Department of Physics and Geology \\
              Perugia University and INFN, Perugia\\
              via A. Pascoli snc 06123 Perugia\\
              \email{sara.fucini@pg.infn.it}           
\and
M. Rinaldi \at
              Department of Physics and Geology \\
              Perugia University and INFN, Perugia\\
              via A. Pascoli snc 06123 Perugia\\
              \email{matteo.rinaldi@pg.infn.it} 
\and
              S. Scopetta \at
              Department of Physics and Geology \\
              Perugia University and INFN, Perugia\\
              via A. Pascoli snc 06123 Perugia\\
              \email{sergio.scopetta@pg.infn.it}  
              }

\date{Received: date / Accepted: date}

\maketitle

\begin{abstract}
 The measurement of nuclear generalized parton di\-stri\-bu\-tions (G\-P\-D\-s)
in hard exclusive processes, such as deeply virtual Compton Scattering
(DVCS), will be one of the main achievements of a new generation
of experiments at high luminosity, such as those under way
at the Jefferson Laboratory  (JLab) with the 12 GeV electron beam and, above
all, those planned at the future Electron Ion Collider.
The CLAS collaboration at JLab has recently demonstrated the possibility
to disentangle the two different channels of nuclear DVCS, the coherent and incoherent
ones, a first step towards
the measurement of GPDs of nuclei and of bound nucleons, respectively,
opening new exciting perspectives in the field. In this scenario,
theoretical precise calculations, ultimately realistic, become mandatory.
Light nuclei, for which realistic studies are affordable and
conventional nuclear effects can be safely estimated, so that possible
exotic effects can be exposed, play an important role.
The status of the calculation of GPDs for light nuclei will be
summarized, in particular for $^3$He and $^4$He, and some updates
will be presented.
The prospects for the next years, related to the
new series of measurements at future facilities,
will be addressed.

\end{abstract}

\section{Introduction}
\label{intro}
The nucleus is a unique laboratory for fundamental studies of the QCD hadron 
structure
\cite{Cloet:2019mql}
. For example, the extraction of the neutron information from light 
nuclei, essential for a precise flavor separation of parton distributions 
(PDFs), the measurement of nuclear PDFs, relevant for the analysis of 
nucleus-nucleus scattering aimed, e.g., at producing quark-gluon plasma, 
or the 
phenomenon of in-medium fragmentation, mandatory to unveil the dynamics of 
hadronization, require nuclear targets. Nevertheless, inclusive 
Deep Inelastic 
Scattering (DIS) experiments off nuclei have been unable to answer a few 
fundamental questions. Among those of interest in this paper, we list: (i) the quantitative microscopic 
explanation of the so called European Muon Collaboration (EMC) 
effect~\cite{Aubert:1983xm}, i.e., the
observed medium 
modification of the nucleon parton structure; (ii) the correct extraction of the structure of the neutron from nuclear data.

Novel coincidence measurements at high luminosity facilities, such as 
Jefferson Laboratory (JLab), have become recently possible, 
addressing a new era in the 
knowledge of the parton structure of nuclei~\cite{Dupre:2015jha}.
An even more relevant improvement is expected
with the operation of the future Electron-Ion Collider (EIC),
based in the US \cite{Accardi:2012qut}.
In particular, two promising directions beyond 
inclusive measurements, aimed at unveiling the three dimensional (3D) 
structure of the bound nucleon, are deep exclusive processes off nuclei,
and semi-inclusive deep inelastic 
scattering (SIDIS) involving nuclear targets.
In deep exclusive processes, as those here addressed, one accesses the 3D structure 
in coordinate space, in terms
of generalized parton distributions (GPDs).
In the following, we will show how, in this way, a relevant contribution 
is expected to the 
solution of long standing problems, such as: (i) the non nucleonic contribution
to nuclear structure, (ii) the quantitative explanation of the 
medium modification of the nucleon parton structure, or (iii) a precise flavor 
separation of the nucleon parton distributions.

The paper is structured as follows. The next section is dedicated to show in general which novel information one could obtain from the measurements of nuclear GPDs. In section 3,
the role of few-body nuclear systems is illustrated. Known results are summarized
and some updates are presented for $^3$He
and $^4$He nuclei. The perspectives in this field of research are eventually described.

\section{Learning from nuclear GPDs}

Excellent introduction to GPDs, quantities initially defined in refs.~ \cite{Mueller:1998fv, Ji:1996ek, Radyushkin:1996nd} and measured in hard exclusive processes, can be found in
refs.~ \cite{Goeke:2001tz,Diehl:2002he,Belitsky:2005qn,Boffi:2007yc}.
Among their most relevant properties,
we remind just the ones of interest in the present paper.
To fix the ideas, let us think to deeply virtual Compton scattering (DVCS), depicted in Fig. 1 in the handbag approximation.
\begin{figure}[t]
\center
\resizebox{0.45\textwidth}{!}
{\includegraphics{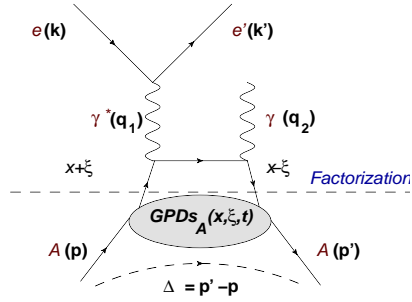}}
\caption{The handbag contribution to the coherent DVCS process
off a nucleus $A$.}
\label{fig:1}       
\end{figure}
If the initial photon virtuality, $Q^2 = -q_1^2 = -(k-k')^2$,
is much larger than $-t = -\Delta^2 = -(p-p')^2 $, the momentum
transferred to the hadronic system
with initial (final) 4-momentum $p(p')$,
the hard vertex of the ``handbag'' 
diagram shown in Fig. 1
can be studied perturbatively. 
The soft part is parametrized in terms of the GPDs
of the target,
which depend on $t=\Delta^2$, on
the so-called skewness  $\xi =-{\Delta^+}/{P}^+$,
i.e., the difference in plus momentum fraction between the initial and the final states, and  on $x$,
the average plus momentum fraction of the struck parton 
with respect to the total momentum, not experimentally accessible
(the notation $a^\pm = (a_0 \pm a_3)/\sqrt{2}$ is used). The average photon four momentum is $q=(q_1 +q_2)/2$,
while the total target momentum is $P = p+p'$.
The number of leading-twist GPDs for a target $A$
with spin $J_A$ is $2(2 J_a+1)^2$, one half of which are chiral odd.
In this paper we will deal only with chiral even
ones, i.e., for example, for a spin 1/2 target,
the two spin-independent ones, defined in terms of an
off forward quark light-cone correlator:
\begin{eqnarray}
\label{eq1}
& & {1 \over 2} \int {d \xi^- \over 2 \pi} e^{i x P^+\xi^- }
\langle P' s' | \, \bar \psi_q \left(0,\xi^-/2,0_\perp\right)
\gamma^+ \, \psi_q \left( 0, \xi^-/2,0_\perp \right) | P s \rangle  =  
\nonumber
\\
& & H_q(x,\xi,\Delta^2) {1 \over 2 }\bar u(P',s') 
\gamma^+ u(P,s) + 
E_q(x,\xi,\Delta^2) {1 \over 2} \bar u(P',s') 
{i \sigma^{+ \nu} \Delta_\nu \over 2M} u(P,s) \, , \quad 
\end{eqnarray}
and the spin dependent ones, $\tilde H_q$
and $\tilde E_q$, obtained from the above expression
substituting $\gamma^+$ and $\sigma^{+ \nu} \Delta_\nu$ 
by $\gamma^+\gamma_5$ and $\gamma_5 \Delta^+$, respectively.

The $x-$integral of the GPDs yields proper
form factors (ffs). For example, $H_q$ and $E_q$ are related
to the Dirac and Pauli ones, respectively.
More in general, Lorentz invariance determines
the polynomiality property,
stating that the $n-$moments of GPDs
are polynomials in $\xi$ of order $n+1$.

The forward limit, $\Delta^2 \rightarrow 0$, of the GPDs $H_q$
and $\tilde H_q$, provides the helicity-independent and helicity-dependent PDFs, respectively.

Initially introduced as a tool towards
the understanding of the spin content of the proton,
being $H_q$ and $E_q$ related to the total angular
momentum through the Ji's sum rule
\cite{Ji:1996ek}, another
crucial property of GPDs is the possibility to allow
the so-called tomography
of the target. As demonstrated in ref.~ \cite{Burkardt:2000za}, 
the GPD $H_q$
at zero skewness, Fourier transformed to the impact parameter space,
represents the number density of the partons with a given longitudinal momentum fraction at a given transverse distance
from the center of the target.
Obviously, having at hand tomography data for the proton, for the nucleus and for the bound proton, possible differences in the distributions of the partons in the transverse plane between these three systems would provide a pictorial representation of the realization of the EMC effect. 
In principle, the tomography of the nucleus and that of the bound proton
are accessible through the so called coherent DVCS, where the nucleus
is detected intact in the final state, and incoherent DVCS, with
the detection of a bound proton in the final state.
Nuclear DVCS is therefore very promising in this direction.

The study of DVCS has therefore many virtues but it is very difficult.
Hard exclusive processes have obviously very small cross sections
and in the final state one has at the same time
slow and fast systems to be detected, which
makes analyses very involved.
On the top of that, the average momentum fraction $x$ is not accessible
and the DVCS amplitude is integrated on this variable, being proportional
to the so-called Compton form factors (CFFs), discussed in the following
sections. 
One should notice that in the inclusive case, i.e., in DIS, the differential cross section is proportional to the PDFs, which makes their measurements much easier
than that of GPDs, which have to be deconvoluted from the CFFs.

Despite of these drawbacks, a first glance at the free proton tomography, 
obtained with little model dependence,
has been obtained~\cite{Dupre:2017hfs}.
The nuclear one, after the
important achievement of the CLAS collaboration at JLab,
with the first separation of coherent and incoherent channels in DVCS off $^4$He \cite{Hattawy:2017woc,Hattawy:2018liu},
should be accessed in the next years.

The relevance of the coherent channel has been stressed in \cite{Dupre:2015jha}.
Here the main advantages are just listed. Besides the already mentioned
nuclear tomography, along the lines of \cite{Burkardt:2000za}, one could
1) compare data with  precise impulse approximation (IA) calculations, possible for few body nuclei, to expose non nucleonic d.o.f., according to an idea initially presented in 
\cite{Berger:2001zb};
2) access  information on the nuclear energy momentum tensor (EMT) and the $d$-term (initial idea in \cite{Polyakov:2002yz}, more recent developments in \cite{Jung:2014jja}, and a 
report in \cite{Polyakov:2018zvc}); we observe that
a positron beam, whose use at JLab and at the EIC is under discussion
\cite{Accardi:2020swt}, would help to access this information; 3) access gluon gpds in nuclei, 
haunting possible gluon d.o.f. \cite{Armstrong:2017wfw}; 4) extract 
pieces of information for the free neutron, possible using specific nuclei
and specific polarization setups in the experiments \cite{Rinaldi:2012ft};
5) study a predicted peculiar shadowing at low x \cite{Goeke:2009tu}.

Two of these relevant possibilities, beyond tomography,
are presented in the two following subsections.

\subsection{Non-nucleonic degrees of freedom in nuclei from nuclear GPDs}
\label{sec:1}

The knowledge of 
GPDs would permit the investigation of the interplay of nucleon and parton 
degrees of freedom in the nuclear wave function
\cite{Berger:2001zb}. In standard DIS off a nucleus with four-momentum $P_A$ and $A$ nucleons of mass $M$, this information can be 
accessed in the region where $x_{Bj} = \frac{Q^2}{2 M \nu}$ is greater 
than 1, $\nu$ being the energy transfer in the laboratory system. In this region, kinematically forbidden for a free proton, extremely fast quarks are tested and measurements are therefore very 
difficult, because of vanishing cross-sections. As explained 
in~\cite{Berger:2001zb,Cano:2003ju}, a similar information can be accessed in 
DVCS at lower values of $x_{Bj}$.

To understand this aspect, it is instructive to analyze coherent DVCS in 
IA. Let us think to unpolarized 
DVCS off a nucleus with $A$ nucleons, which is dominated by the GPD $H_q^A$. This case has been treated in \cite{Cano:2003ju} for the deuteron target, 
in~\cite{Guzey:2003jh} for spin-0 nuclei, in~\cite{Kirchner:2003wt} for nuclei 
with spin up to 1, in~\cite{Scopetta:2004kj} for $^3$He and in~\cite{Liuti:2005gi,Fucini:2018gso} for $^4$He.  
In IA, $H_q^A$ is obtained as a 
convolution between the non-diagonal spectral function of the internal 
nucleons and the GPD $H_q^N$ of the nucleons themselves. 
The process is depicted in Fig.~\ref{fig:1} for coherent 
DVCS, assuming the handbag approximation. Here,
one parton  belonging to 
a given nucleon interacts with the probe and it is then 
reabsorbed, with an additional momentum $\Delta$, by the same nucleon, 
without further re-scattering with the recoiling system. 
Finally, the interacting nucleon is reabsorbed back 
into the nucleus. The main assumptions
of IA are: (i) the nuclear operator is 
just the sum of single nucleon free operators, i.e., there are only 
explicit
nucleonic d.o.f.; (ii) the interaction of the debris originating 
by the struck nucleon with the remnant (A - 1) system is disregarded    
(we are close to the Bjorken limit); (iii) the coupling of the virtual photon with the 
spectator (A-1) system is neglected (given the high momentum transfer), (iv) 
the effect of the boosts is not considered (they can be properly taken into 
account in a light-front framework). 
Whitin all these conditions, the GPD
 $H_{q}^A$ can be 
written in the form:
\begin{eqnarray}
H_{q}^A(x,\xi,\Delta^2) =  
\sum_N \int_x^1 \frac{dz}{z}
h_N^A(z, \xi ,\Delta^2 ) 
H_q^N \left( \frac{x}{z},
\frac{\xi}{z},\Delta^2 \right)
\label{main}
\end{eqnarray}
where
\begin{eqnarray}
h_N^A(z, \xi ,\Delta^2 ) =  
\int d E \int d \vec p \, P_N^A(\vec p, \vec p + \vec \Delta,E) 
\delta \left( z + \xi  - \frac{p^+}{P^+} \right)
\label{hq0}
\end{eqnarray}
is the off-diagonal light-cone momentum distribution of the nucleon $N$ in the 
nucleus $A$. 
$P_{N}^A (\vec p, 
\vec p + \vec \Delta, E )$ is the one-body off-diagonal spectral function. The latter quantity has been 
introduced and calculated for the 
$^3$He target
first in~\cite{Scopetta:2004kj}. Here,  $E=E_{min} + E^*_R$ is the so called removal energy, with 
$E_{min}=| E_{A}| - | E_{A-1}|$ and $E^*_R$ is the excitation energy of the 
nuclear recoiling system.

One should notice that eq.~(\ref{main}) fulfills the general properties of 
GPDs, i.e., the forward limit reproduces the standard nuclear 
PDF in IA and the the first $x$-moment yields the IA form factor. The polynomiality property is formally fulfilled,  however, since in these calculations use has been made of non-relativistic 
wave functions, this feature is actually valid only at order $O( {p^2 \over m^2} )$. 
In particular,
by taking the forward limit ($\Delta^2 \rightarrow 0, \xi \rightarrow 0$) of 
eq.~(\ref{main}), one gets the expression which is usually found, for the 
parton distribution $q_A(x)$, in the IA analysis of unpolarized DIS:
\begin{eqnarray}
q_A(x_{Bj}) & = &  H_q^A(x_{Bj},0,0) 
 =  \sum_{N} \int_{x_{Bj}/A}^1 { d \tilde z \over \tilde z}
f_{N}^A(\tilde z) \, q_{N}\left( {x_{Bj} \over \tilde z}\right).
\label{mainf}
\end{eqnarray}
In the latter equation,
\begin{eqnarray}
f_{N}^A(\tilde z) & = & h_{N}^A(\tilde z, 0 ,0) 
=  \int d E \int d \vec p \, P_{N}^A(\vec p,E) 
\delta\left( \tilde z - { p^+ \over P^+ } \right)
\label{hq0f}
\end{eqnarray}
is the light-cone momentum distribution of the nucleon $N$ in the nucleus, 
$q_N(x)= H_q^N( x, 0, 0)$ is the distribution of the quarks of 
flavor $q$ in the nucleon $N$ and $P_N^A(\vec p, E)$ is the one body diagonal 
spectral function.

In a typical IA calculation, the light-cone momentum distribution $f_{N}^A(z)$ 
turns out to be strongly peaked around the value $z \simeq 1/A$.
Therefore, in order to investigate the effects of nucleons with large ``plus'' momentum fraction, one needs to consider
 regions where
therefore to be at $z > 1/A$. Looking at the lower integration limit in 
eq.~(\ref{mainf}), it is clear that, in the DIS case, this occurs at 
$x_{Bj} > 1$, where the cross sections are very small. 
Recent analyses of inclusive data 
at $x_{Bj} > 1$ have only been able to quantify 
the 
number of such fast correlated nucleons, but without addressing their internal 
structure~\cite{Egiyan:2005hs,Subedi:2008zz,Hen:2012yva}.

In the coherent channel of a hard exclusive process,
when the recoiling nucleus does not breakup,
a longitudinal momentum is transferred,
tuned by the kinematical variable $\xi$,
which can be larger than the narrow width of the light cone
momentum distribution, so that the predicted 
DVCS cross section is vanishing.
In~\cite{Cano:2003ju}, for the 
deuteron case, it is estimated that the IA predicts a vanishing cross section 
already for $ x_{Bj} \simeq 0.2$, i.e. for $\xi \simeq 0.1$. By experimentally 
tuning $\xi$ in a coherent DVCS process one could therefore explore at 
relatively low values of $x_{Bj}$ contributions to the GPDs not included in 
IA, i.e., accessing non nucleonic degrees of freedom generating correlations at parton 
level or other exotic effects. These studies
could contribute to explain the nuclear EMC effect.

\subsection{Spatial distribution of energy, momentum and forces experienced by 
partons in nuclei}
\label{sec:2}

In this section, we shall discuss how the lowest Mellin moments of GPDs 
provide us with information about the spatial distribution of energy, momentum 
and forces experienced by quarks and gluons inside nuclei. This idea, leading 
to predictions to be experimentally tested, has been developed initially 
in~\cite{Polyakov:2002yz}. To be specific, let us consider a spin-$1/2$ hadronic 
target, e.g. a nucleon. All spin independent equations apply to the spin-$0$ 
targets as well.

The $x$-moments of the GPDs are related to three scalar ffs~\cite{Ji:1996ek}, in terms of which
the matrix elements of the EMT can be para\-me\-trized.
The ff we are 
interested in, usually called $d^Q(t)$, is related to the first 
Mellin moment of the unpolarized GPD~\cite{Ji:1996ek}:

\begin{eqnarray}
\label{jisr} 
\int_{-1}^1 dx\ x\ H(x,\xi,t)=M_2^Q(t) +\frac 45\ d^Q(t)\ \xi^2\, . 
\end{eqnarray}
Thanks to this relation, $d^Q(t)$ can be studied in hard exclusive processes. 
In particular, $d^Q(t)$ contributes with an $x_{Bj}$ independent term to the 
real part of the DVCS amplitude, which is accessible through the beam charge 
asymmetry \cite{Brodsky:1972vv}, measurable using
electron and positron beams. At the same time,  this ff is related to the 
so-called D-term in the parametrization of the GPDs \cite{Polyakov:2002wz}. 
Let us remind that
at 
small $x_{Bj}$, $t$ and to the leading order in $\alpha_s(Q)$, the $x_{Bj}$ 
dependent contribution to the real part of the DVCS amplitude is basically 
given by the ``slice" $H_q(\xi,\xi,t)$ of quark GPD, measurable through
the DVCS beam spin asymmetry (BSA)
\begin{equation}\label{alu}
    A_{LU}= \frac{d\sigma^+ - d\sigma^-}{d \sigma^+ + d\sigma^-} \, ,
\end{equation}
where the differential cross sections $d \sigma^\pm$ for the beam helicities ($\pm$) appear.

In principle, the ff $d^Q(t)$ can be therefore 
extracted from combined data of DVCS beam spin asymmetry and beam charge 
asymmetry, whose measurements requires positron beams (see,
e.g., ref.~ \cite{Accardi:2020swt} for the pespectives in the future use of positrons).
It can be shown that $d^Q(t)$ is related 
to the traceless part of $T_{ik}^Q(\vec r,\vec s)$, which characterizes the 
spatial distribution (averaged over time) of shear forces experienced by 
quarks in the nucleon~\cite{Polyakov:2002wz}.
In ref.~\cite{Polyakov:2002wz}, to illustrate the physics of $d^Q(t)$, a 
simple model of a large nucleus has been considered. In this framework, as an example, use has been made of   
a liquid drop model for a nucleus, with sharp edges. In this case
one gets $d(0)\sim A^{7/3}$, i.e. it 
rapidly grows with the atomic number. This fact implies that the contribution 
of the D-term to the real part of the DVCS amplitude grows with the atomic 
number as $A^{4/3}$. This should be compared to the behavior of the amplitude 
$\sim A$ in IA and experimentally checked by measuring the charge beam 
asymmetry in coherent DVCS on nuclear targets. A similar $A$ dependence of 
$d(0)$ has been predicted also in a microscopic evaluation of nuclear GPDs for 
spin-0 nuclei in the framework of the Walecka model~\cite{Guzey:2005ba}. The 
meson (non-nucleonic) degrees of freedom were found to strongly influence DVCS 
nuclear observables, in the HERMES kinematics.

The first experimental study of DVCS on nuclei of noble gases, reported 
in~\cite{Airapetian:2009cga}, was not able to observe the predicted $A$ 
dependence. The data are anyway affected by sizable error bars and more 
precise DVCS experiments could provide information on nuclear modifications of the 
EMT ffs. 

\section{GPDs of few body nuclei}

Since conventional nuclear effects, if not properly evaluated, can be easily 
mistaken for exotic ones, light nuclei, for which realistic calculations are 
possible and conventional nuclear effects can be exactly calculated, play a 
special role. Besides, light nuclei impose their relevance in 
the extraction of the neutron information, necessary to perform a clean flavor 
separation of GPDs, crucial to test QCD fundamental symmetries and 
predictions. We note that an indirect procedure to constrain the neutron GPDs 
using coherent and incoherent DVCS off nuclei has been proposed 
in ref.~ \cite{Guzey:2008th}.

\subsection{The deuteron}

The deuteron, being a spin 1 nucleus,
has a very rich spin structure and in particular, as detailed in
the introduction, 18 leading twist GPDs.
It offers therefore unique possibilities.  However, since the deuteron  is
 very weakly bound, it is not expected to be suitable to investigate  nuclear medium modifications.
It is the first natural candidate to
access the neutron information, using mainly incoherent DVCS.
In that channel,  FSI could hinder their extraction, and
specific kinematical regions, where FSI is known
to be less relevant, have to be therefore selected. To this aim,
dedicated theoretical estimates of FSI in this channel
are very important. An experiment
of this kind has been performed at JLab 
after the 12 GeV upgrade and the analysis is in progress~\cite{silvia}.
Another promising possibility, for the measurement
of DVCS off the neutron, to be detailed in forth-coming
proposals~\cite{Hafidi:2009loi}, is that offered by the detection
of a slow recoiling proton in DVCS off the deuteron. In this case,  the experimental setup, successfully
used in spectator SIDIS by the BONUS
collaboration at JLab~\cite{KKW}, will be exploited.
Among the nuclear systems,
the deuteron obviously represents
the easiest to teat theoretically.
A
realistic nuclear description and a light-front (LF) relativistic implementation of IA are available.
We recall that the twist-2 deuteron GPDs have been first introduced in
in ref.~ \cite{Berger:2001zb}.  Then,  in ref.~ \cite{Cano:2003ju}, the latter quantities
have been evaluated in IA on the light cone together with the related coherent DVCS
cross section. In addition, 
a specific spin sum rule has been discussed in ref.~ \cite{Taneja:2011sy}.
The deuteron will not be further discussed  in this paper.
The reader interested in recent developments
can consult the discussion of
transversity GPDs in ref.~ \cite{Cosyn:2018rdm}; 
the fulfillment of polinomiality in calculations of deuteron GPDs in ref.~
\cite{Cosyn:2018thq}); the EMT for the deuteron in ref.~
\cite{Cosyn:2019aio}.

\subsection{$^3$He}

Trinucleons, having spin and isospin 1/2 as the nucleon,
are suitable to access  isospin and flavor dependence of 
nuclear effects
\cite{Scopetta:2004kj,Scopetta:2009sn}. Here we discuss the 
advantages offered by trinucleons
targets and beams.
For these light nuclei
realistic IA calculations are available and the off-diagonal
spin-dependent spectral function has been calculated 
\cite{Scopetta:2004kj,Scopetta:2009sn,Rinaldi:2012ft,Rinaldi:2012pj,
Rinaldi:2014bba}.
Moreover, a relativistic treatment of the above quantity  has 
been developed, for the moment only in the forward case, in ref.~ \cite{DelDotto:2016vkh}. Let us remark 
that relativistic effects could be relevant in the calculations of 
PDFs and related quantities.
In addition, the specific spin structure of  $^3$He
allows to extract the neutron spin-dependent GPDs, by  properly
taking into account conventional nuclear effects.

\begin{figure}[t]
\center
\vspace{-2cm}
\hspace{-2cm}
\includegraphics[width=8cm]{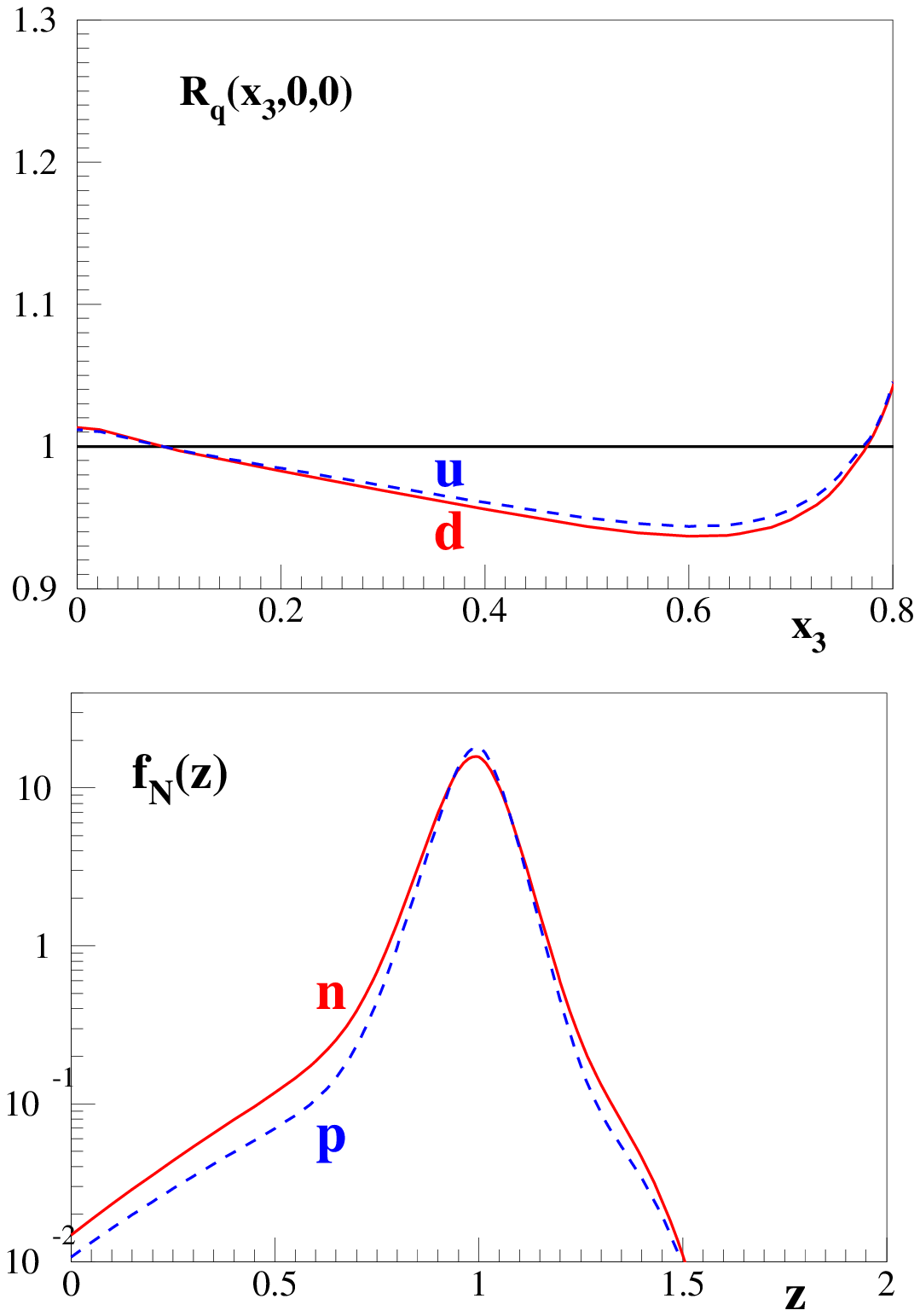}
\hspace{-3cm}
\includegraphics[width=8cm]{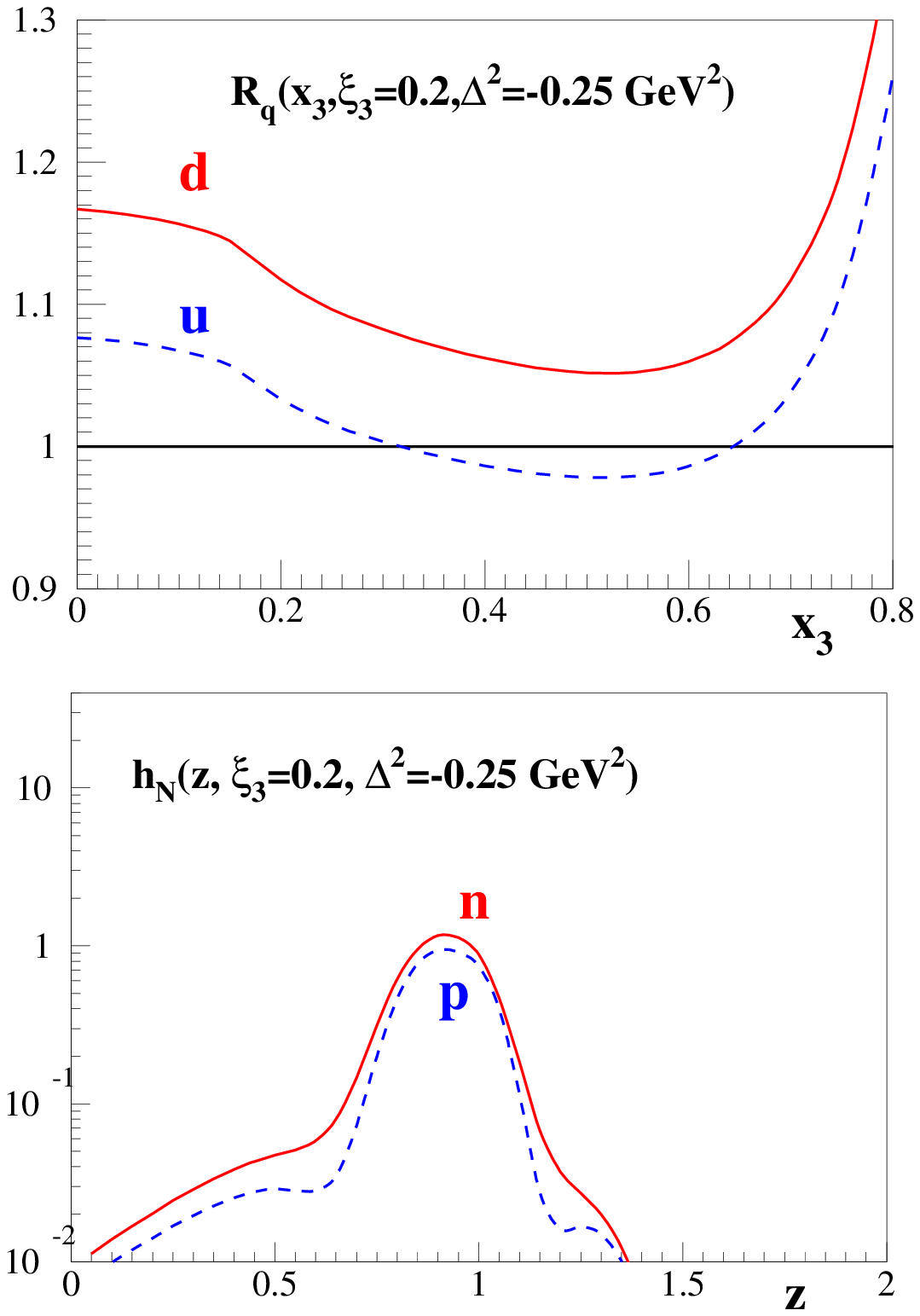}
\vspace{-3cm} 
\caption{Upper left panel: the dashed (full) line represents the ratio of the 
$^3$He GPD $H$ to the corresponding quantity of the constituent nucleons (2 
protons and one neutron), for the u (d) flavor, in the forward limit, as a 
function of $x_3=3x$. Lower left panel: the dashed (full) line represents the light 
cone momentum distribution, eq.~(\ref{hq0f}), for the proton (neutron) in 
$^3$He.
Upper right panel: the same as in the upper left panel, 
but at off-forward kinematics: $t = -0.25$ GeV$^2$ and $\xi_3 = 3 \xi = 0.2$. 
Lower right panel: the dashed (full) line represents the light cone off diagonal 
momentum distribution, eq.~(\ref{hq0}), for the proton (neutron) in $^3$He at 
the same off-forward kinematics.
}
\label{fig:lcforward}       
\end{figure}

As  shown in section 1, the conventional treatment of
nuclear GPDs, 
through the IA, involves a non-diagonal nuclear spectral function. Despite
its complicated 
dependence on the momentum and removal energy, 
the latter 
can be exactly 
evaluated  for $^3$He.
Therefore, such a nucleus is very suitable to realistically study 
 polarization
effects and to attempt a precise flavor 
separation of GPDs, being not an isoscalar target. 

A realistic microscopic calculation of the unpolarized quark GPD of the $^3$He 
nucleus has been presented in ref.~\cite{Scopetta:2004kj}. 
The proposed scheme 
points to the coherent channel of hard exclusive processes. Nuclear effects, 
evaluated within the AV18 potential~\cite{Wiringa:1994wb}, are found to be 
larger than in the forward case and get bigger and bigger with 
increasing $t~(\xi)$ and keeping 
$\xi~(t)$ fixed. Besides, the 
obtained GPD 
cannot be factorized into  $t$-dependent and  $t$-independent terms, 
as 
suggested by prescriptions proposed for finite nuclei
\cite{Kirchner:2003wt}.

In ref.~\cite{Scopetta:2009sn} the analysis has been extended, showing that other 
conventional nuclear effects, such as isospin and binding ones, or the 
uncertainty related to the use of a given nucleon-nucleon potential, are 
rather bigger than in the forward case. An example is shown in 
Fig.~\ref{fig:lcforward}. As one can see, these 
effects 
increase when the light-cone momentum distributions, eqs.~(\ref{hq0}) 
and~(\ref{hq0f}), depart from a delta-like behaviour, as
it happens in the non-forward case. In addition, nuclear 
effects, 
for the $u$ ($d$) flavor, follow the path of the proton (neutron) light-cone 
momentum distributions. 
Future experimental observations of this behaviour, a 
typical 
prediction of a realistic conventional IA approach, would provide 
relevant 
information on the reaction mechanism of DIS off nuclear targets. Let 
us point out that these kind of effects could not be observed 
in  isoscalar targets, such as $^2$H or $^4$He.

\begin{figure}[h]
\center
\resizebox{0.8\textwidth}{!}{\includegraphics{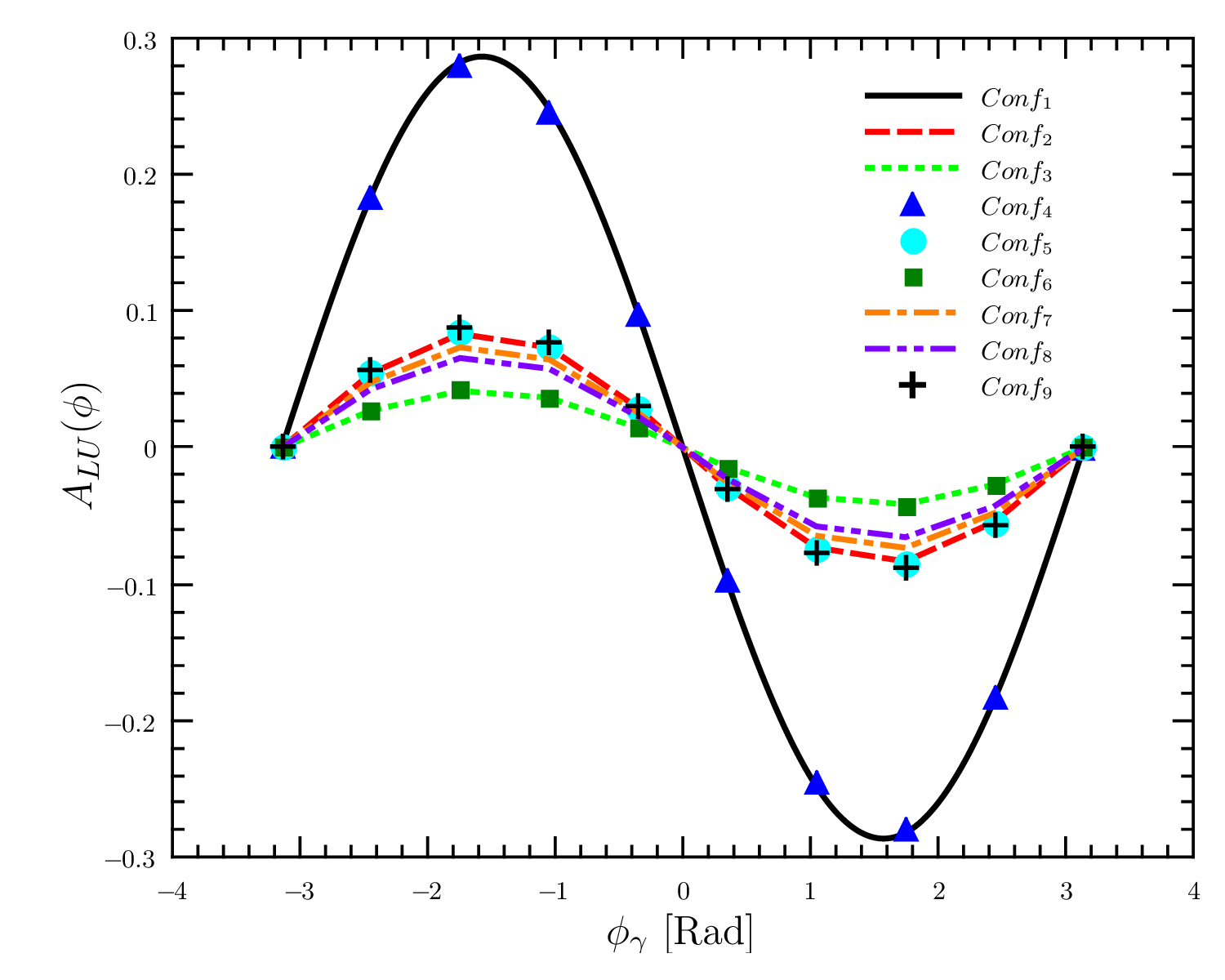}}
     \caption{The preliminary $^3$He beam-spin asymmetry evaluated
for the 9 kinematical  configurations given in Tab. \ref{tab3}. Here $\phi_\gamma=\phi-\pi$, with
$\phi$ the azimuthal angle between the hadron and lepton planes.}
\label{asy3}       
\end{figure}

In ref.~\cite{Rinaldi:2012pj} the $E$ GPD of $^3$He has been 
calculated following the line of ref.~ \cite{Scopetta:2004kj}. Let us 
remark that the GPDs $H$ and $E$ are involved in the Ji's sum 
rule
which allows 
to extract  information on the 
parton angular momentum 
of the system. One of the main outcomes of the analysis is that the neutron 
contribution is the dominant one to the GPD $E$ and in particular to 
the combination $H+E$. We recall that the  $^3$He magnetic form 
factor can be obtained as the first moment of 
$H+E$ and, moreover, such a combinations enters the Ji's sum rule. Therefore, these results confirmed, also in the case of DVCS, that $^3$He is an ideal 
target to access properties related to the polarization of the neutron. 
In fact, a polarized $^3$He nucleus is, to an extent of 90\%,  equivalent to a polarized 
neutron.
Furthermore, in ref.~ \cite{Rinaldi:2012ft}, a technique, able to take 
into account  the nuclear 
effects included in an IA analysis, and to safely extract the neutron 
information, has been proposed. In particular it has been shown that 
such a procedure works in the relevant kinematic conditions  of  possible 
measurements.
A similar technique has been successfully tested also for 
the extraction 
of the  $\tilde H$ GPD of the neutron from the corresponding quantity 
of $^3$He~\cite{Rinaldi:2014bba}. This investigation 
would require 
coherent DVCS off polarized $^3$He, a challenging but not impossible 
measurement at present facilities. 
Thanks to these observations, coherent DVCS should be considered as 
a golden process to access the neutron GPDs and, in turn, the 
total angular 
momentum of the quarks in the neutron. 
Indeed, the $E$ GPD of 
isoscalar 
targets, such as $^2$H and $^4$He, is very small and, therefore, they are not suitable for this purpose.
However, the 
measurement of this
GPD would  requires a  transverse polarization of $^3$He, a very 
difficult setup for the coherent channel at the present 
facilities. Nevertheless,
in order to guide future experimental analyses, at JLab
and at the future EIC, we are at present using the calculated
GPDs to estimate the actual observables to be measured in experiments. As a preliminary outcome of these studies,
we show in Fig. \ref{asy3}
the beam-spin asymmetry  for the 
coherent DVCS processes off $^3$He, calculated in the 
kinematical configurations listed in 
Tab. \ref{tab3}, similar to those alreastudied at JLab 
using $^4$He and discussed in the next section.
Let us point out that
together with the AV18 spectral function previously discussed,  use has been made of the nucleon GPDs of ref.~ \cite{Goloskokov:2007nt} to evaluate the nuclear GPDs.
 Predictions for the 
relative 
cross-section will be available soon, also in the EIC kinematics.

\begin{table}
\begin{center}
\begin{tabular}{|c|c|c|c|c|}
\hline
$x_B$& $Q^2$ &$t$   & $\xi$ & Configuration 
\\
& [GeV$^2$] &[GeV$^2$]   &  &   
\\\hline
0.132 & 1.18 & -0.095 & 0.070664 & 1   \\\hline
0.171 & 1.45 & -0.099 & 0.093494 & 2  \\\hline
0.225 & 1.84 & -0.106 & 0.126761 & 3   \\\hline
0.134 & 1.15 & -0.096 & 0.071811 & 4 \\\hline
0.171 & 1.42 & -0.099 & 0.093494 & 5 \\\hline
0.223 & 1.87 & -0.106 & 0.125492 & 6 \\\hline
0.159 & 1.37 & -0.081 & 0.086366 & 7  \\\hline
0.179 & 1.51 & -0.094 & 0.098298 & 8 \\\hline
0.193 & 1.61 & -0.126 & 0.106807 & 9 \\\hline
\end{tabular}
\end{center}
\caption{The kinematical configurations adopted  to evaluate the beam-spin 
asymmetry.}
\label{tab3}
\end{table}

In closing thi  let us stress that
$^3$He represents a unique system for nuclear GPD studies:   
its conventional structure is completely under control, so that the interplay of conventional and exotic effects can be exposed  to provide 
hints on them, when heavier nuclear targets are used.
Besides, polarized $^3$He is a convenient
effective polarized neutron whose GPDs $E$ and $\tilde H$ could be easily extracted  from the corresponding $^3$He quantities, at low 
$t$, in a rather
model independent way. To this aim, the
measurements of polarized
coherent DVCS would be required, 
certainly challenging but, in particular for the extraction of $\tilde H$, unique and not 
prohibitive. This is true especially at the future EIC,
where polarized light ion beams are planned to circulate.

\subsection{$^4$He}

In $^4$He,
the binding energy per nucleon is significantly larger than that of the other light nuclei so that such nucleus exhibits the features of a typical nuclear system. For this reason, $^4$He can be considered as a prototype system to study the partonic structure of typical nuclear targets. Besides, the study of such nucleus exhibits many advantages: from one side, being spinless, it has a straightforward theoretical description in terms of one leading-twist GPD in the chiral even sector and realistic (non relativistic) descriptions, although challenging, can be developed. From the other side, the
experimental analyses are easier with respect to other targets like the $^3$He nucleus, as discussed in the previous section. In facts,  the two DVCS channels, the coherent and the incoherent ones, have been  recently disentangled at JLab by the CLAS collaboration \cite{Hattawy:2017woc,Hattawy:2018liu}.  Several theoretical models have been developed to explain such reaction mechanisms
\cite{Guzey:2003jh,Liuti:2004hd}. The most recent calculations within, an Av18 \cite{Wiringa:1994wb} + UIX \cite{Pudliner:1995wk} semi-realistic nuclear description, has been presented in ref.~ \cite{Fucini:2018gso} for the coherent channel,  and in refs.~ \cite{Fucini:2019xlc,Fucini:2020lxi}
for the incoherent one. These models turn out to be able to successfully reproduce experimental data for $A_{LU}$ in the kinematical range unraveled at JLab, i.e. the valence region. A deeply investigation of the region at small x$_{Bj}$ is going on as these effects could be sizeable and useful for the explanation of the EMC effect in the shadowing region, that could be probed by the forth-coming EIC. Besides, for their features, these models have been used to develop a new Monte Carlo event generator for the DVCS process off $^4$He in view of the big experimental program that will be carried out at the EIC. In the Wiki page \cite{Wiki}, the main results from the generation of the events for the coherent channel are shown.

Let us present our main results for these processes.
As stated before in this paper,
the $x$ dependence of GPDs cannot be experimentally accessed. Nevertheless,
  the so called Compton Form Factor (CFF) $\mathcal{H}$ can be extracted from measured observables. As a matter of facts, GPDs ($H_q)$ are hidden in the imaginary and the real part of such objects defined ($e_q$ being the quark electric charge) as:\begin{equation}\label{imeq}  \Im m {\cal H}(\xi,t)=\sum_{q} e_q^2[ \, H_q^{^4He}(\xi,\xi,\Delta^2)- H_q^{^4He}(-\xi,\xi,\Delta^2) \, ]\,,
\end{equation}
\begin{equation}
	\Re e {\cal H}(\xi,t)= \Pr \sum_{q}e_q^2 \int_{-1}^{1} dx \bigg(\frac{1}{\xi-x}-\frac{1}{\xi+x}\bigg)
	H_q^{^4He}(x,\xi,t) \, ,
\nonumber
\label{reeq}
	\end{equation}
	respectively. The experimental observable which gives access to the above quantities is the beam spin asymmetry (BSA), defined in eq.~ (\ref{alu}).
	
 In the following, we will summarize the main results obtained from a realistic calculations of conventional effects for the BSA within a plane wave impulse approximation approach.

\subsubsection{Coherent DVCS channel} 

\begin{figure}
    \centering
    \includegraphics[angle=0,scale=0.5]{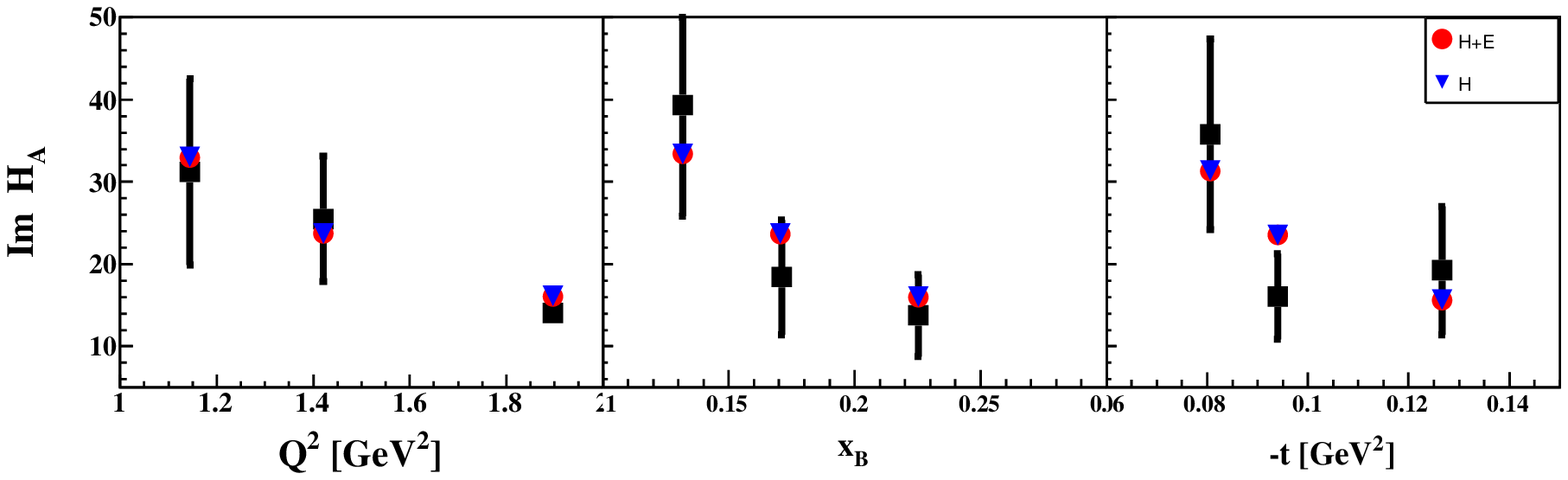}
    \includegraphics[angle=0,scale=0.5]{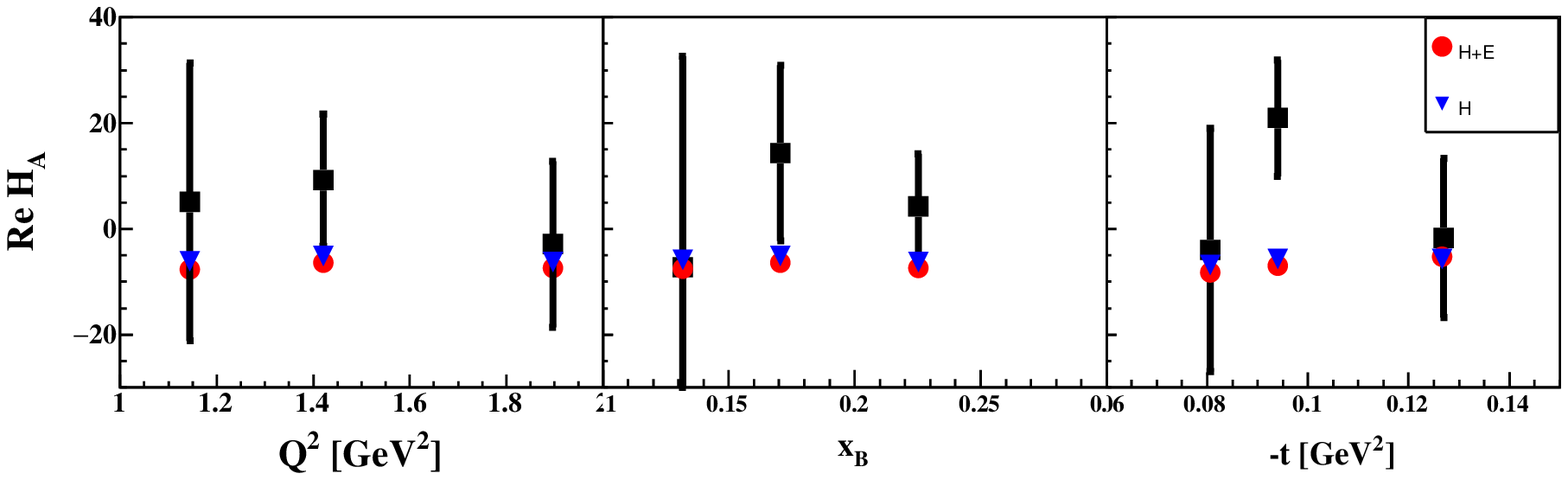}
    \caption{Imaginary (upper panel) and real part (lower panel) of the CFF for  $^4$He: results of ref.~\cite{Fucini:2018gso} (blue triangles) and new results including also the GPD E (red dots) compared with data (black squares) \cite{Hattawy:2017woc}. The quantity
is shown in the experimental $Q^2$, $x_B=Q^2/(2 M \nu)$ and $t=\Delta^2$ bins, respectively.}
    \label{imcoh}
\end{figure}

\begin{figure}
\center
   \includegraphics[scale=0.5]{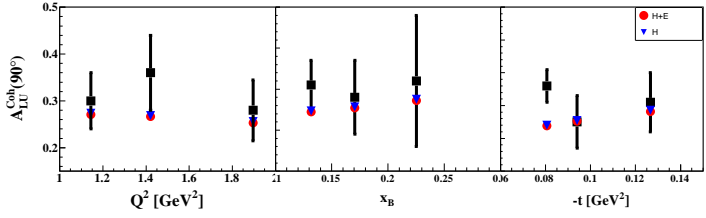}
    \caption{$^4$He azimuthal beam-spin asymmetry
$A_{LU}(\phi=90^o)$: results of ref.~ \cite{Fucini:2018gso} (blue triangles) and
new results including also the GPD E$^q_N$ (red dots), compared with data (black squares) \cite{Hattawy:2017woc}. From left to right, the quantity
is shown in the experimental $Q^2$,  $x_B=Q^2/(2M\nu)$ and $t=\Delta^2$ bins, respectively}
    \label{alucoerente}
\end{figure}

The most general coherent DVCS process $A(e,e'\gamma)A$ allows to study the partonic structure of the recoiling whole nucleus $A$ through the formalism of GPDs. In the IA scenario presented above, a workable expression for the quantity $H_q^{^4He}(x, \xi,\Delta^2)$, the GPD of the quark of flavor q in the $^4$He nucleus, is obtained as a convolution between the GPDs $H_q^N$of the quark of flavor $q$ in the bound nucleon N and the off-diagonal light-cone momentum distribution of N in $^4$He and reads

\begin{equation}
H_q^{^4He}(x,\xi, \Delta^2)= \sum_N \int_{|x|}^1 { dz \over z } 
	h_N^{^4He}(z,\xi,\Delta^2)
\boldsymbol{H}_q^N\bigg(\frac{x}{z},\frac{\xi}{z},\Delta^2\bigg)\,.
\label{gpd}
\end{equation}
In the previous equation, $\boldsymbol{H}_q^N = \sqrt{1-\xi^2}[H_q^N -\frac{\xi^2}{1-\xi^2}E_q^N]$ has been evaluated using the GPD model given in refs.~ \cite{Goloskokov:2007nt,Goloskokov:2008ib}. 
Given in the kinematical range accessible at JLab, the contribution of the GPD E$_q^N$ is supposed to be negligible and 
in ref.~\cite{Fucini:2018gso}, only the contribution of the GPD H$_q^N$ has been taken into account.
 Here, for the first time,
 both the  H$_q^N$ and the E$_q^N$ GPD contributions
 have been accounted, as it can be seen in Figs.  \ref{imcoh}-  \ref{alucoerente}.
Then, the light cone momentum distribution in Eq (\ref{gpd}) is defined as \begin{equation}
h_N^{^4He}(z,\Delta^2,\xi)= 
\int dE \, \int d \vec p \,
P^{^4He}_N(\vec p, \vec p + \vec \Delta, E) \delta \bigg(z - \frac{\bar p^+}{ \bar P^+}\bigg)\,,
\end{equation}
where the off diagonal spectral function $P_N^{^4He}(\vec p, \vec p + \vec \Delta,E)$
represents the probability amplitude to have a nucleon with momentum $\vec p$ that leaves the nucleus, generating an excited recoiling system with energy $E_R^*=E-|E_A|+|E_{A-1}|$, with $|E_A|$ and $|E_{A-1}|$ the nuclear binding energies, and then goes back with a momentum transfer $\vec \Delta$.
In order to have a full realistic evaluation of $P_N^{^4He}$, an exact
description of all the $^4$He spectrum, including three-body scattering states, is needed. Though, it represents a demanding few body problem whose complete evaluation is still going on. 
Old results in the diagonal case can be found in
refs. \cite{Morita:1991ka,Efros:1998eb}.
For this reason, as an intermediate step, in the present calculation a model of the nuclear non-diagonal spectral function \cite{Viviani:2001wu}, based on the momentum distribution corresponding
to the Av18 NN interaction ref.~\cite{Wiringa:1994wb} and including 3-body forces \cite{Pudliner:1995wk}, has been used for the excited 3- and 4- body states.
For the ground state, exact wave functions of 3- and 4-body systems
\cite{Kievsky:1992um,Viviani:2004vf}, evaluated along the scheme of ref.~ \cite{Kievsky:2008es}, have been considered. Before
showing the
 comparison with data, we remark that checks for the nuclear charge form factor and for nuclear parton distributions have been successfully performed. Then, the comparison of our model for $H_q^{^4He}$ with the measured observables has been achieved evaluating numerically eqs.~ (\ref{imeq},\ref{reeq}). The obtained results are depicted in Figs. \ref{imcoh}.
Finally, also the BSA of the coherent DVCS, i.e.
\begin{equation}
A_{LU}(\phi) = 
\frac{\alpha_{0}(\phi) \, \Im m(\mathcal{H}_{A})}
{\alpha_{1}(\phi) + \alpha_{2}(\phi) \, \Re e(\mathcal{H}_{A}) 
+ \alpha_{3}(\phi) 
\bigg( \Re e(\mathcal{H}_{A})^{2} + \Im m(\mathcal{H}_{A})^{2} \bigg)} \, ,
\nonumber
\end{equation}
 where $\alpha_i(\phi)$ are kinematical coefficients  defined in ref.~ \cite{Kirchner:2003wt}, has been calculated and compared with the experimental data.
As shown in Fig. \ref{alucoerente}, a very good agreement is found with the data. These results lead us to conclude that a careful analysis of the reaction mechanism in terms of basic conventional ingredients is successful at the present experimental accuracy without requiring the use of exotic arguments, such as dynamical off-shellness.

\subsubsection{Incoherent DVCS channel}
\begin{figure}[h]
\hspace{-0.8cm}
    \includegraphics[scale=0.28]{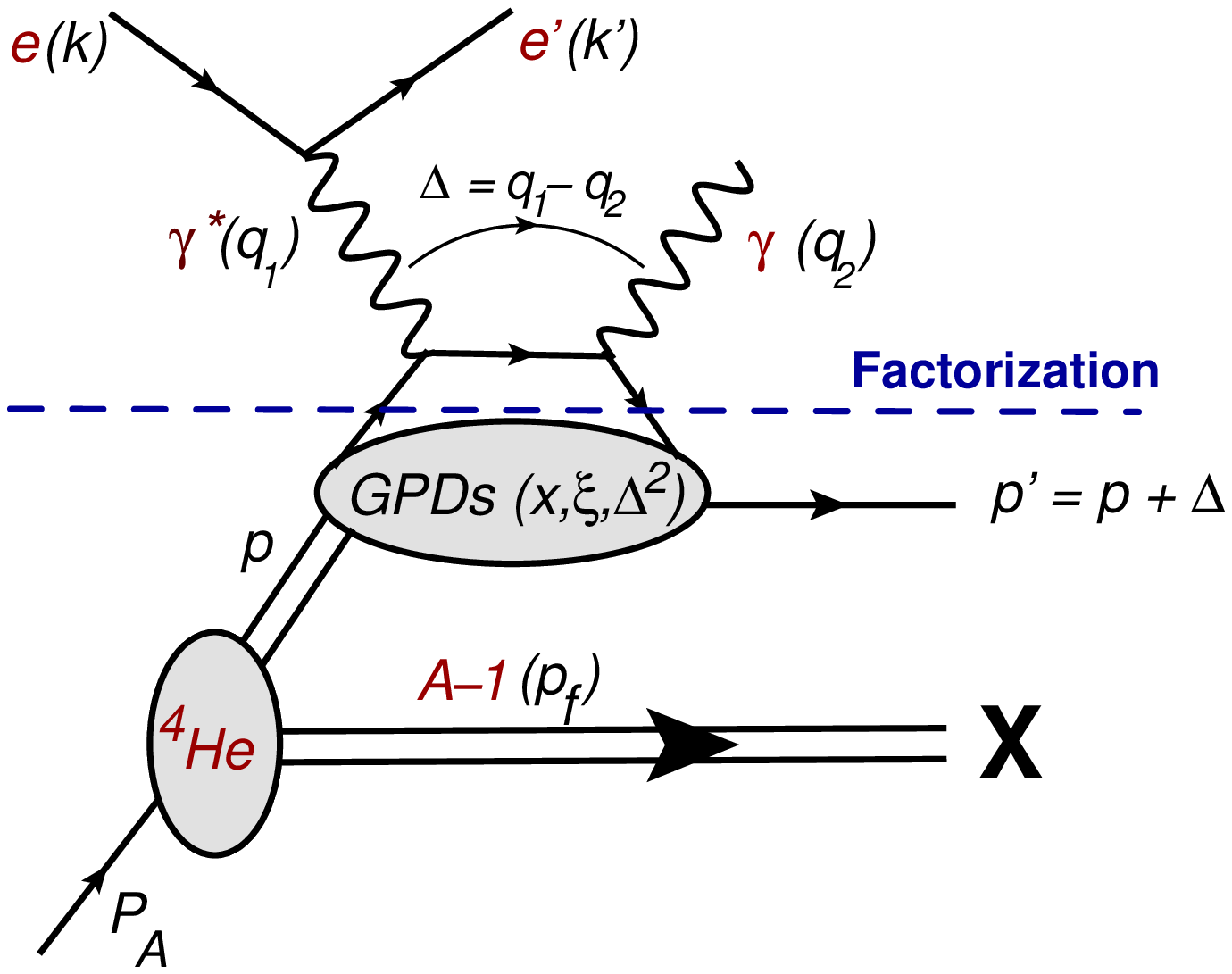}
    \hspace{0.5cm}
    \includegraphics[scale=0.28]{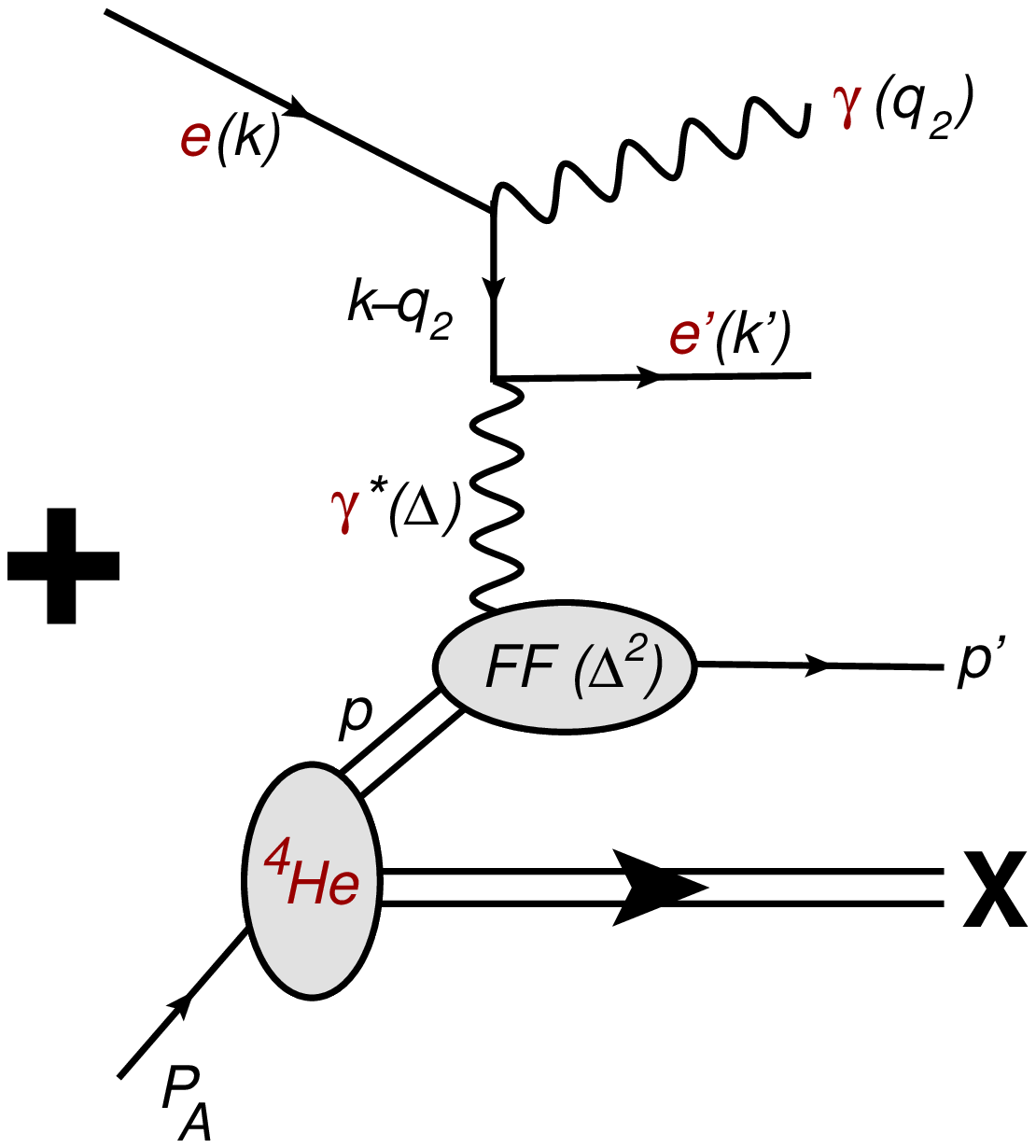}
    \hspace{0.5cm}
     \includegraphics[scale=0.28]{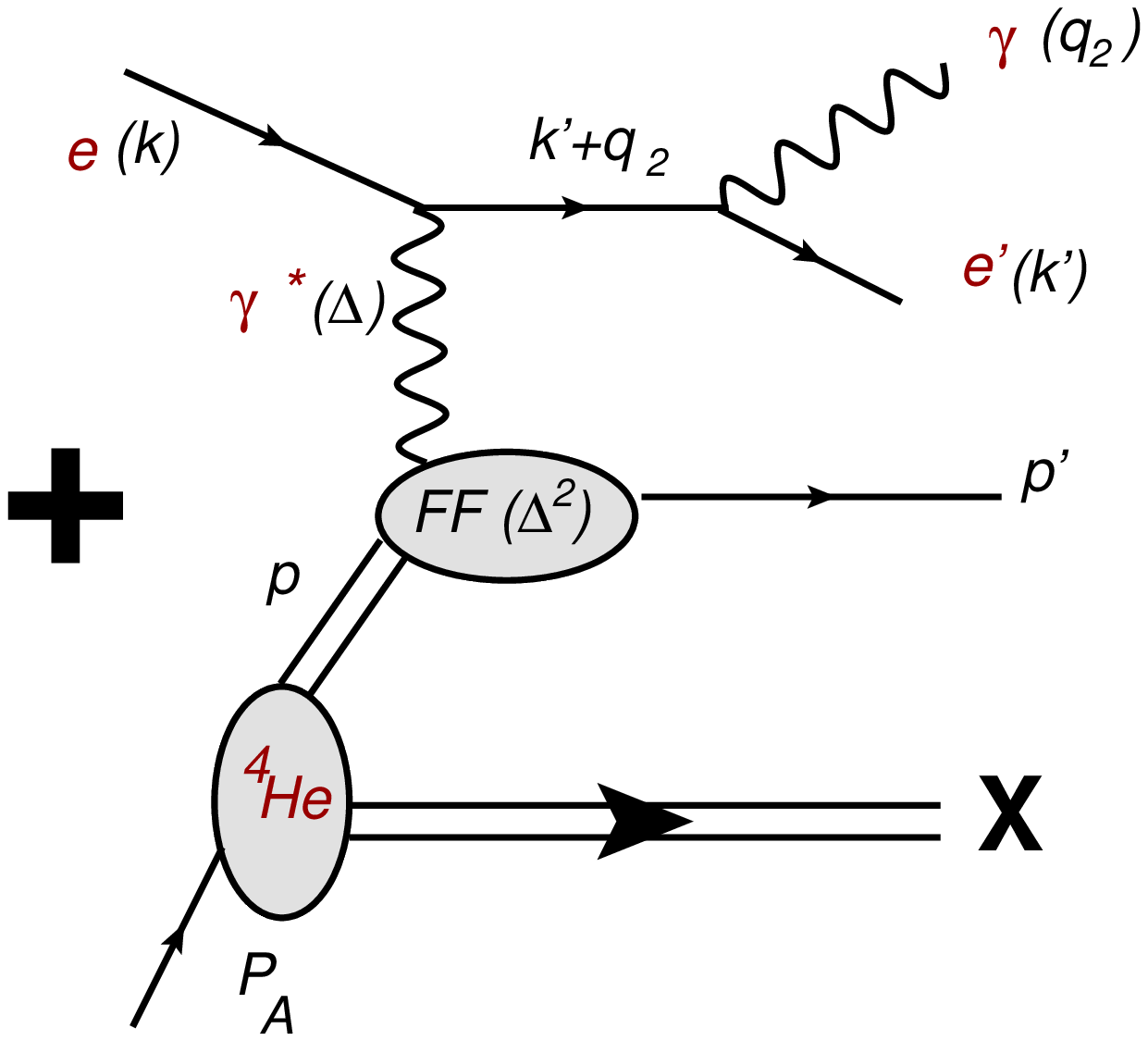}
    \caption{Incoherent DVCS off $^4$He in IA. To the left, pure DVCS contribution; to the right the two Bethe Heitler terms.}
    \label{incodvcs}
\end{figure}
In the process $A(e,e'\gamma p)X$  depicted in Fig. \ref{incodvcs}, the parton structure of the kicked out bound proton can be studied. For a complete evaluation of eq. (\ref{alu}), the expression of the cross-section for a DVCS process off a bound proton in $^4$He is required. In the IA scenario, only the kinematical off-shellness of the initial bound proton has been accounted for. In this way, a convolution formula for the differential cross has been derived, and it reads
\begin{equation}
d \sigma^\pm \equiv  \frac{d\sigma^\pm_{Inc} }{d{x}_{Bj} dQ^2 d{\Delta}^2 d\phi}= \int_{exp} dE \, d{\vec p} \,P^{^4He}(\vec {p},E)
|\mathcal{A}^{\pm}({\vec p}, E ,K)|^2
g(\vec{p},E,K) \nonumber\, .
\end{equation}
In the equation above,
$K$ is the set of kinematical variables $\{x_{Bj},Q^2,t,\phi\}$ probed in the experiment that selects only the relevant part of the diagonal spectral function $P_N^{^4He}(\vec p, E)$, which has therefore to be integrated only in the selected experimental range $exp$.
Moreover, the quantity $ g(\vec{p},E,K)$ is a complicated function, whose derivation is detailed in ref.~ \cite{Fucini:2020lxi}.
Finally, in the above equation, the squared amplitude includes three different terms, i.e  $\mathcal{A}^2= T_{DVCS}^2+T_{BH}^2+\mathcal{I}_{DVCS-BH}$ as shown in Fig. \ref{incodvcs} and each contribution has to be evaluated for an initially moving proton.
\begin{figure}
\centering
    \includegraphics[scale=0.5]{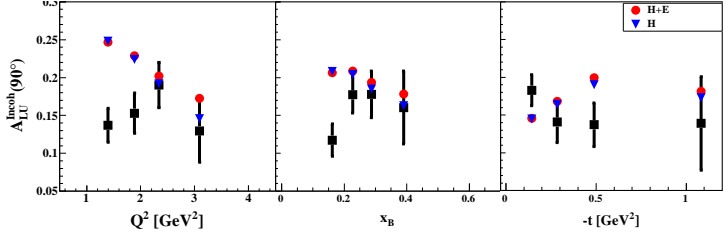}
    \caption{Azimuthal beam-spin asymmetry for the proton in
$^4$He, $A_{LU}^{Incoh}$,
eq.~ \ref{aluratio},
for $\phi = 90^o$: results presented
in ref.~ \cite{Fucini:2019xlc}(red dots), results updated
including also the contribution of the GPD $E$ (blue triangles), compared with data
(black squares) \cite{Hattawy:2018liu}. Data and our results are represented in the experimental bins, from left to right, for Q$^2$, $x_B=Q^2/(2M\nu)$ and $t=\Delta^2$.}
    \label{aluinco}
\end{figure}
The amplitude addressed in refs.~ \cite{Fucini:2019xlc,Fucini:2020lxi}
 generalize those obtained for a proton at rest ( see ref.~ \cite{Belitsky:2005qn}).  Details on the main assumptions here adopted are summarized in refs.~ \cite{Fucini:2019xlc,Fucini:2020lxi}. 
Since in the kinematical range probed at JLab, the DVCS contribution is smaller than the BH term, information on the partonic structure of the target can be inferred only through the interference DVCS-BH term. In this way, the BSA reads
\begin{equation}\label{aluratio}
A_{LU}^{Incoh} = \frac{
\int_{exp} dE \, d \vec p \, 
{ {P^{^4He}(\vec p, E )}} 
\, g(\vec p,E,K)\, 
{{\mathcal{I}_{DVCS-BH}}} 
}
{
\int_{exp} dE \, d \vec p \, {{P^{^4He} (\vec p, E )}} 
\, g(\vec p, E,K)\,
{{T_{BH}^2}}
}\equiv \frac{\mathcal{I}_{DVCS-BH}^{^4He}}{T_{BH}^{2\, ^4He}}
\end{equation}
For an exhaustive comparison with the experimental data, the azimuthal dependence of eq.~\ref{aluratio}, given by the decomposition in $\phi$ harmonics of the interference and the BH part, has been exploited. Information about the parton content of the bound proton enters only the interference term, i.e. $I_{DVCS-BH}\propto \Im m \,\mathcal{\vec H}_q(\xi',t)$. This term accounts for the modification to the partonic structure through the skewness $\xi' = \frac{-q^2}{P\cdot q}$, where the off-shellness and the dependence on the 3-momentum components of the bound proton, affected by nuclear effects, enter. In the CFF $\vec H_q(\xi,t)$, contributions from the GPDs H and E, for which we made use of the GK models \cite{Goloskokov:2007nt, Goloskokov:2008ib}, has been included in the present results. As for the coherent channel, the contribution from the GPD E, not already included in refs.~ \cite{Fucini:2019xlc, Fucini:2020lxi}, turns out to be very small at the considered kinematics.
The results are depicted in Fig. \ref{aluinco}. As expected, the agreement with experimental data is good except the region of lowest $Q^2$ where the impulse approximation exposes its limitations and FSI effects could be considerable. In order to estimate 
 how nuclear effects affect the obtained results, i.e. if they are related to some medium modification of the inner parton structure described by the GPD, we considered the ratio between the BSA for a bound nucleon, given by eq. ~\ref{aluratio} and that for a free proton (labelled with the superscript \textit{p}):
\begin{equation}
\frac{ A_{LU}^{Incoh} } { A_{LU}^p } 
\propto \frac{ \mathcal{I}_{DVCS-BH}^{^4He} } { \mathcal{I}_{DVCS-BH}^{p} }
\frac{ T_{BH}^{2\, \, p} } { T_{BH}^{2\, \, ^4He} }= \frac
{(nucl.eff.)_{Int}}{(nucl.eff.)_{BH}}\,.
\label{aluratio1}
\end{equation}
As a matter of facts, the above \textit{super ratio} is proportional to the ratio of the nuclear effects on the BH and DVCS interference to the nuclear effects on the BH cross section.
If the nuclear dynamics modifies the $\mathcal{I}_{DVCS-BH}^{^4He}$ and $T_{BH}^{2\, ^4He }$ cross sections in a different way, the effect
can be big even if the parton structure
of the bound proton does not change appreciably.
However, such effects could be dramatically reduced and hidden in the \textit{super ratio}.\\ This is what we definitely obtained: within our IA approach, both in the BH and in the interference amplitudes big nuclear effects occur that are cancelled out in the \textit{super ratio}. Such effects seem to be related to kinematical nuclear effects, foreseen in a conventional nuclear physics scenario, rather than to medium modifications of the parton structure due exotic effects, such as dynamical off-shellness
 More detailed studies in this sense are presented in ref.~ \cite{Fucini:2020lxi} where also checks for the stability of our model, using different ingredients both for the nuclear part and the nucleonic GPDs models, have been performed. Concerning this latter ingredient as well as the relevant GPD E$_q^N$, use has been made of the virtual access infrastructure \texttt{3DPARTONS}~ \cite{Berthou:2015oaw}.

\section{Perspectives}

Since the time of the first measurements of nuclear GPDs, performed
by the HERMES collaboration in DESY~%
\cite{Airapetian:2009cga}, which
could not differentiate coherent and incoherent channels directly 
and had to rely on the dominance of either channel at small and large $t$,
a crucial step forward has been performed by 
the data released by the CLAS collaboration at JLab,
successful in distinguishing coherent and incoherent DVCS channels exclusively~%
\cite{Voutier:2013gia,Hattawy:2017woc,Hattawy:2018liu}. 
The analysis clearly shows only a small coverage in $x_{Bj}$ and $t$ 
and we should expect that an extension of this program with the upgraded CLAS12 
will provide a large data set to analyze the GPDs of light nuclei in the valence 
region, with much smaller errorbars, given the luminosity gain. Farther in the future, the EIC in the 
US~\cite{Accardi:2012qut} will be the perfect facility  to study nuclear DVCS. 
Indeed, thanks to the collider kinematics, it will be much easier to 
detect the recoiling nuclei and to polarize the incoming nuclear beams. In addition, 
 the high energy available atthe  EIC
 will allow
 to cleanly map the nuclear GPDs at low $x$, including also the gluon GPDs.  
Besides, as for any coherent exclusive process, in DVCS off nuclei the collider setup will make easy the detection of the recoiling intact nucleus,
which is very slow in a fixed target experiment. This is even more important in the incoherent channel, for which the detection of other nuclear fragments could allow to control FSI effects.

The study of nuclear GPDs is a fascinating subject, already deeply studied theoretically, as described
in this paper. At the same time, its actual experimental investigation has just started and,
in the near future, exciting updates to the material presented here are expected. With the operation of light nuclear beams at the EIC,
the involvement of the Few-Body Physics community will become more and more important.

\begin{acknowledgements}
We gratefully acknowledge the collaboration with R. Dupr\'e and M. Viviani,
and many crucial discussions with E. Voutier and M. Hattawy on the Jefferson Lab esperiments.
S.F. thanks P. Sznajder and C. Mezrag for some tuition on the
use of the virtual access infrastructure 3DPARTONS, funded by the European Union’s Horizon 2020 research and innovation programme under grant agreement No 824093.
This work was supported in part by the STRONG-2020 project of the European Union’s Horizon 2020
research and innovation programme under grant agreement No 824093, Working Package 23, "GPDS-ACT" and by
the project ``Deeply Virtual Compton Scattering off $^4$He", in the programme FRB of the University
of Perugia.
\end{acknowledgements}



\printbibliography


%
%

\end{document}